\documentclass[12pt,epsfig,epsf]{article}

\usepackage{cite}
\usepackage{slashed}
\usepackage{epsf,color,colordvi}
\usepackage{graphicx}
\usepackage{amsmath}
\usepackage{amssymb}
\usepackage{epsfig}
\usepackage{cite}
\usepackage{fontenc}
\usepackage{float}
\usepackage[lofdepth,lotdepth,caption=false]{subfig}
\usepackage{color}
\usepackage{booktabs}
\usepackage{tabularx}
\usepackage{multirow}
\usepackage{longtable}
\usepackage{lscape}
\usepackage{dcolumn}
\usepackage{rotating}
\usepackage[english]{babel}
\usepackage[autostyle]{csquotes}
\usepackage{hyperref}
\usepackage{xcolor}
\usepackage[normalem]{ulem}

\def\hlpm#1{\textcolor{black}{\textrm{#1}}}

\newcommand{\bi}{\begin{itemize}}
\newcommand{\ei}{\end{itemize}}

\def\beq{\begin{equation}}
\def\eeq{\end{equation}}

\newcommand{\bea}{\begin{eqnarray}}
\newcommand{\eea}{\end{eqnarray}}

\newcommand{\beqa}{\begin{eqnarray}}
\newcommand{\eeqa}{\end{eqnarray}}

\newcommand{\ldm}{\Delta m_{31}^2}
\newcommand{\sdm}{\Delta m_{21}^2}

\newcommand{\ie}{{\it i.e.}}

\newcommand\schd{Schr$\ddot{\rm o}$dinger}

\def\lsim{\mathrel{\rlap{\lower4pt\hbox{\hskip1pt$\sim$}}
    \raise1pt\hbox{$<$}}}         
\def\gsim{\mathrel{\rlap{\lower4pt\hbox{\hskip1pt$\sim$}}
    \raise1pt\hbox{$>$}}}         

\newcommand{\nue}{\ensuremath{\nu_e}}
\newcommand{\numu}{\ensuremath{\nu_\mu}}

\begin{document}

\begin{titlepage}

\renewcommand{\thefootnote}{\alph{footnote}}

\vspace*{1.cm}

\begin{flushright}
\end{flushright}

\renewcommand{\thefootnote}{\fnsymbol{footnote}}
\setcounter{footnote}{0}

{\begin{center}
{\large\bf
Enhanced violation of Leggett-Garg Inequality in three flavour neutrino  oscillations via non-standard interactions \\
[0.2cm]
}
\end{center}}

\renewcommand{\thefootnote}{\alph{footnote}}

\vspace*{.8cm}
\vspace*{.3cm}
{
\begin{center} 
            {\sf 
                Sheeba Shafaq$^\S$\,\footnote[1]
               {\makebox[1.cm]{Email:} sheebakhawaja7@gmail.com} and 
                Poonam Mehta$^{\S}$\,\footnote[2]{\makebox[1.cm]{Email:}
                pm@jnu.ac.in}
}
\end{center}
}
\vspace*{0cm}
{\it 
                                                                                               \begin{center}
$^\S$\, School of Physical Sciences,\\   Jawaharlal Nehru University, 
      New Delhi 110067, India

\end{center}}

\vspace*{1.5cm}


{\Large 
\bf
 \begin{center} Abstract  
\end{center} 
 }
 
\vspace*{.5cm}

Neutrino oscillations occur due to non-zero masses and mixings and  most importantly they are believed to maintain quantum coherence even over astrophysical length scales. In the present study, we explore the quantumness of three flavour neutrino oscillations by studying the extent of violation of Leggett-Garg inequalities (LGI) if non-standard interactions are taken into account. We  report an enhancement in violation of LGI with respect to the standard scenario for appropriate choice of NSI parameters. 

\end{titlepage}

\section{Introduction}

Even though quantum mechanics was born in nineteen twenties~\cite{diracbook}, several significant, conceptual and foundational developments which stem from quantum mechanics emerged much later. The Aharonov-Bohm effect was understood in the sixties~\cite{PhysRev.115.485}, Bell's inequalities~\cite{RevModPhys.38.447} and  the issue of entanglement were appreciated in seventies and developments related to the Leggett-Garg inequalities (LGI)~\cite{Leggett:1985zz} emerged in the eighties. 
 In their seminal paper, Leggett and Garg~\cite{Leggett:1985zz} derived a class of inequalities which provided a way to test the applicability of quantum mechanics as we go from the microscopic to the macroscopic world. 
 The work was based on our intuition about the macroscopic world which can be defined in terms of the two principles  :  
(a) Macroscopic realism (MR) which implies that the performance of a measurement on a macroscopic system reveals a well-defined pre-existing value 
 (b) Non-invasive measurability (NIM) which states that in principle, we can measure this value without disturbing the system.  
The classical world, in general, respects both these assumptions. However, in quantum mechanics, both the assumptions are violated as it is based on superposition principle
 and  collapse of wave function under measurement (see~\cite{emary2013leggett} for a review).
 There exists a  maximum quantum-mechanical value of the LGI correlator  with the standard measurement protocol and this is referred to as Luder's bound or temporal Tsirelson's bounds~\cite{Cirelson:1980ry,PhysRevLett.111.020403,PhysRevLett.113.050401,Rohrlich:1995cf}. {{LGIs can be used as indicators of quantum coherence, specifically for macroscopic systems.  Apart from LGI, there are other useful measures which prove useful to quantify quantumness such as quantum witness, contextuality etc~\cite{PhysRevLett.113.140401,PhysRevLett.116.120404,Ban2019,PhysRevA.95.022101,Mendoza-Arenas2019}.}}

If we look at the developments in the neutrino sector, soon after the discovery of the second type of neutrino in the sixties, the idea of neutrino flavour oscillations was proposed~\cite{mns,Ponte1,Ponte2,Gribov:1968kq}. The experimental vindication of the idea of neutrino flavour oscillations took several decades and was rewarded with the 2015 Nobel Prize for Physics~\cite{nobel2015}. Neutrino oscillations among the three active flavours imply that at least two of the neutrino states are massive which can not be reconciled within the Standard Model of particle physics.  The phenomenon of neutrino flavour oscillation arises from the phase difference acquired by the mass eigenstates due to their time evolution during propagation in vacuum or matter~\cite{giunti}. The idea of non-standard interactions (NSI) originated in the seminal work of Wolfenstein as a viable alternative to mass-induced oscillations~\cite{PhysRevD.17.2369} (see also~\cite{Wolfenstein:1979ni,Valle:1987gv, Guzzo:1991cp, Roulet:1991sm}). Of course, now we know that standard  mass-induced neutrino oscillations have been firmly established yet sub-dominant effects due to NSI can impact measurements at current and future oscillation experiments. There are strong reasons to believe that sub-dominant effects due to neutrino NSI could cause interference with the standard oscillation measurements and neutrino  NSI is the most studied new physics topic in the current times (see~\cite{Ohlsson:2012kf,Farzan:2017xzy,Dev:2019anc} for reviews).  
 Recent studies have demonstrated that the discrepancy arising in data from two of the long baseline neutrino experiments, T2K and NoVA can be reconciled by invoking large NSI~\cite{Chatterjee:2020kkm,Denton:2020uda}.

 Given that neutrinos exhibit sustained quantum coherence even over astrophysical length scales, it is natural to explore geometric aspects of the phases involved~\cite{Mehta:2009ea}  as well as  think about quantification of the coherence properties of neutrinos via temporal correlations in the form of LGI. Study of temporal correlations in the form of LGI has attracted significant attention in recent times both in the context of two~\cite{Gangopadhyay:2013aha,Formaggio:2016cuh,Fu:2017hky} and three~\cite{Gangopadhyay:2017nsn,Naikoo:2017fos,Naikoo:2019eec,Naikoo:2019gme} flavour neutrino oscillations.  It should be noted that while different dichotomic observables have been employed in these studies, the  neutrino  matter interactions  have been considered to be standard in these studies.

Before we proceed to describe the key idea of the present work, we would like to summarize the existing work on LGI in the context of neutrino oscillations.  In one of the early  attempts~\cite{Gangopadhyay:2013aha}, the authors considered LGI in the context of two state oscillations of neutral kaons and neutrinos.  The implication of neutrino oscillations on LGI was characterized by a quantity  $K_4$  which was found to be sensitive to the mixing angle appearing in two flavour neutrino oscillations and the conclusion was that it could reach    its upper bound  of $2\sqrt2$ for a specific value of the mixing angle, $\theta=\pi/4$.  Observation of violation of LGI was reported for the first time in the context of Main Injector Neutrino Oscillation Search (MINOS) experiment~\cite{Formaggio:2016cuh}. The large ($\sim 6\sigma$) violation in this microscopic system of neutrinos over a macroscopic distance of 735 km provided the longest range over which a temporal analogue of Bell's test of quantum mechanics had been performed. Soon after, observation of $\sim 6\sigma$ violation of LGI in the experimental data obtained at Daya Bay  reactor experiment was reported~\cite{Fu:2017hky}. It should be remarked that all the above studies were performed assuming two neutrino states only and therefore obscured the parameter dependencies present in the three flavour analysis. In order to  shed some light on the dependence on CP phase and other parameter dependencies via LGI, it becomes imperative to perform the analysis taking  three flavour neutrino oscillations into account. 
The three flavour analysis was first carried out in~\cite{Gangopadhyay:2017nsn} in which condition for attaining maximum violation of LGI was laid down. Moreover, it was concluded that non-zero three flavour oscillation parameters such as $\theta_{13}$,  the CP phase ($\delta$),  the mass ordering parameter ($\alpha$)  led to an enhanced violation of LGI in comparison to   the two flavour case.  Additionally,  three flavour analysis has been carried out assuming stationarity condition~\cite{Naikoo:2017fos} and relaxing it~\cite{Naikoo:2019eec}. With the stationarity assumption, it was  shown that the quantity representing LGI has a sensitive dependence on the neutrino mass ordering~\cite{Naikoo:2017fos}. Various inequivalent forms of LGI   in subatomic systems have been explored in~\cite{Naikoo:2019gme}. 

Using the tools of quantum resource theory~\cite{PhysRevLett.113.140401,PhysRevLett.116.120404}, the authors of Ref.~\cite{Song:2018bma,Ming:2020nyc}   quantified the quantumness  of experimentally observed neutrino oscillations. The authors in~\cite{Song:2018bma} analysed ensembles of reactor and accelerator neutrinos at distinct energies from a variety of neutrino sources, including Daya Bay (0.5 km and 1.6 km), Kamland (180 km), MINOS (735 km), and T2K (295 km).  Though far-fetched, there was also an idea where it was shown that one could obtain difference in possible violation of LGI depending on the type of neutrinos (Dirac or Majorana) by selecting appropriate quantity for $K_3$~\cite{Richter:2017toa}.   
Among other measures of coherence in the context of neutrino oscillations, contextuality has been studied in~\cite{Richter-Laskowska:2018ikv} and the $l_1$ norm of coherence introduced in~\cite{PhysRevLett.113.140401} have been explored in~\cite{Dixit:2018gjc,Dixit2021}.
The entropic uncertainty relations have been investigated by comparing the experimental observation of neutrino oscillations to predictions in~\cite{Wang:2020vdm}. Also, tri-partite entanglement in neutrino oscillations has been studied in~\cite{Jha:2020hyh}.

 The work of Leggett and Garg~\cite{Leggett:1985zz} is, undoubtedly, one of the most profound developments in the area of foundations of quantum mechanics. The present article weaves together the idea of LGI and neutrino oscillation physics in presence of NSI to explore  the extent and possibility of enhancement of violation of LGI in case of three flavour oscillations.  Such a type of enhancement is of interest to a wide range of physicists in several areas of physics and in particular, in the area of quantum information and computation. To the best of our knowledge, NSI induced effects on violation of LGI in neutrino sector have not been reported so far.

The plan of this article is as follows. In Sec.~\ref{sec:fr}, we  describe the basic  framework which comprises of brief review of the three flavour neutrino oscillations in presence of NSI as well as definition of  observables used to quantify the extent of violation of LGI   in the context of neutrino oscillations. We describe our results in Sec.~\ref{sec:res}. Finally, the conclusion and outlook is presented in Sec.~\ref{sec:con}.

\section{Framework}
\label{sec:fr}

The parameters entering the standard  three flavour oscillation framework are : three angles ($\theta_{12}$, $\theta_{23}$, $\theta_{13}$), one phase ($\delta$) as well as the two mass-squared differences ($\Delta m^2_{21}=m^2_2 - m^2_1$ and $\Delta m^2_{31}=m^2_3 - m^2_1$).   Table~\ref{tab} summarizes the current values of the parameters  (taken from~\cite{deSalas:2020pgw}).

 \begin{table}[h]
\centering
\scalebox{0.9}{
\begin{tabular}{ l  c  c  c }
\hline
&&&\\
Parameter & Best-fit value & 3$\sigma$ interval  \\
&&&\\
\hline
&&&\\
$\theta_{12}$ [$^\circ$]             & 34.3                    &  31.4 - 37.4     \\
$\theta_{13}$  [$^\circ$]  (NO)  & 8.58              &  8.16  -  8.94     \\
$\theta_{13}$  [$^\circ$]  (IO)  & 8.63              &  8.21  -  8.99    \\
$\theta_{23}$   [$^\circ$]  (NO)      & 48.8                     &  41.63  - 51.32      \\
$\theta_{23}$  [$^\circ$]    (IO)    & 48.8                     &  41.88  - 51.30      \\
$\sdm$ [$ 10^{-5} \text{eV}^2$]  & $7.5 $  &  $[6.94 - 8.14]$    \\
$\ldm$   [$10^{-3} \text{eV}^2$] (NO) & $+2.56  $   &  $[2.46 - 2.65] $   \\
$\ldm$  [$10^{-3} \text{eV}^2$] (IO) & $-2.46  $  & $-[2.37 - 2.55] $   \\
$\delta$ [Rad.]   (NO) & $-0.8\pi$   & $[-\pi, 0]  \cup [0.8\pi, \pi]$ \\
$\delta$  [Rad.]  (IO) & $-0.46\pi$   & $[-0.86\pi, -0.1\pi]$    \\
&&&\\
\hline
\end{tabular}}
\caption{\label{tab}
The best-fit and allowed range of the standard oscillation parameters used in our analysis~\cite{deSalas:2020pgw}.  
}
\end{table}
 
\subsection{Non-standard interactions} 
\label{sec:nsi}

Non-standard interactions~\cite{PhysRevD.17.2369,Valle:1987gv,Guzzo:1991cp,Roulet:1991sm} (see~\cite{Ohlsson:2012kf,Farzan:2017xzy,Dev:2019anc} for reviews) refer to a wide class of  new physics scenarios  parameterised in a model-independent way at low energies ($E \ll M_{EW}$, where $M_{EW}$ is the electroweak scale) by using effective four-fermion interactions. 
In general, these NSI can impact the neutrino oscillation signals via  charged current (CC)  or neutral
current (NC) processes.     CC interactions affect processes
only at the source or the detector which are are  
discernible at near detectors while  the NC
interactions affect the propagation of neutrinos which can be probed  at far detectors.  
In this work, we consider the NC terms which affect propagation of neutrinos. 
The effective Lagrangian describing the neutrino NSI    is given by 
\bea
\label{nsilag}
\mathcal{L}_{\text{CC}} &=& -2\sqrt{2}G_F\varepsilon_{\alpha\beta}^{ff'C} [\overline{\nu}_\alpha\gamma^\mu P_L l_\beta] ~ [\overline{f}'\gamma_\mu P_C f]~,
\\
\mathcal{L}_{\text{NC}} &=& -2\sqrt{2}G_F\varepsilon_{\alpha\beta}^{fC}[\overline{\nu}_\alpha\gamma^\mu P_L \nu_\beta] ~ [\overline{f}\gamma_\mu P_C f]~.
\eea
where $G_F$ is the Fermi constant,  $\alpha, \beta$ denote lepton flavours and $f$ is the first\footnote{Matter contains only first generation fermions and hence second or
third generation fermions do not affect oscillation experiments.} generation 
fermion ($e,u,d$).   The dimensionless coefficients, $\varepsilon_{\alpha\beta}^{ff'C}$ or $\varepsilon_{\alpha\beta}^{f C}$ quantify the strength of the NSI with respect to the standard weak interaction. Here $f$ and $f'$ denote the charged fermions involved in the interactions with  background fermions. The chirality projection operators are given by 
  $P_L = (1 - \gamma_5)/2$ and $P_C=(1\pm \gamma
_5)/2$. 
It should be noted that only the incoherent sum of all individual
contributions ($e,u,d$)  impacts the coherent
forward scattering of neutrinos on matter. Normalizing 
 to $n_e$, the effective NSI parameter for neutral Earth matter
 is 
\bea
\varepsilon_{\alpha\beta} &=& \sum_{f=e,u,d} \dfrac{n_f}{n_e}
\varepsilon_{\alpha\beta}^f = \varepsilon_{\alpha\beta}^e +2
\varepsilon_{\alpha\beta}^u + \varepsilon_{\alpha\beta}^d + \dfrac{n_n}{n_e}
(2\varepsilon_{\alpha\beta}^d + \varepsilon_{\alpha\beta}^u) = \varepsilon
^e_{\alpha\beta} + 3 \varepsilon^u_{\alpha \beta} + 3
\varepsilon^d_{\alpha\beta} \ ,
      \label{eps_combin}
\eea 
where $n_f$ is the density of fermion $f$ in medium traversed by the
neutrino and $n$ refers to neutrons.  Also,   NC type NSI matter effects are sensitive only to the
vector sum of NSI couplings, \ie, $\varepsilon_{\alpha\beta}^f=
\varepsilon_{\alpha\beta}^{fL} + \varepsilon_{\alpha\beta}^{fR}$.

As far as  CC NSI terms are concerned, 
 those are tightly constrained~\cite{Davidson:2003ha}.  
  The  constraints on  NC type NSI parameters are less stringent.  As
mentioned above, the combination that
enters oscillation physics is given by
Eq.~(\ref{eps_combin}). The individual NSI terms such as
$\varepsilon_{\alpha \beta}^{f L}$ or $\varepsilon_{\alpha \beta}^{f R}$ are
constrained in any experiment (keeping only one of them non-zero at a
time) and moreover the coupling is either to $e,u,d$
individually~\cite{Davidson:2003ha}. In view of this, it is not so
straightforward to interpret those bounds in terms of {an} effective
$\varepsilon_{\alpha \beta}$. {\hlpm{One could take the conservative approach i.e, use the most stringent constraint on individual NSI terms. Using this, it is found that the NSI parameters involving the muon sector are more tightly constrained than  the electron or tau sector ($|\varepsilon_{\mu\mu}| < 0.003$, $|\varepsilon_{\mu\tau}| < 0.05$, $|\varepsilon_{e\mu}| < 0.05$, $|\varepsilon_{e\tau}| < 0.27$, $|\varepsilon_{e e}| < 0.06$, $|\varepsilon_{\tau\tau}| < 0.16$)~\cite{Davidson:2003ha}. }} 
However, the authors in Ref.~\cite{Biggio:2009nt}  deduced model-independent bounds  (assuming that the errors on individual NSI 
 terms are not correlated) on effective NC NSI terms given by
$
\varepsilon_{\alpha\beta} \lsim \left\{  \sum_{C=L,R} [ (\varepsilon_{\alpha \beta}^{e C} )^2 + (3 \varepsilon_{\alpha\beta}^{u C})^2 + (3 \varepsilon _{\alpha \beta}^{d C})^2 ] \right\}^{1/2} $. 
{\hlpm{For neutral Earth matter,  we have 
 \begin{eqnarray}
 |\varepsilon_{\alpha\beta}|
 \;<\;
  \left( \begin{array}{ccc}
4.2  &
0.33 & 
3.0 \\
0.33 &0.068 & 0.33 \\
3.0  &
0.33 &
21 \\
  \end{array} \right) \ .\nonumber
  \label{largensi}
\end{eqnarray} }}
The  values of NSI parameters considered in this work are well within these constraints.

In the ultra-relativistic limit, the neutrino propagation is governed
by a \schd-type equation  with an effective
Hamiltonian
\bea 
{\mathcal
H}^{}_{\mathrm{}} &=& {\mathcal
H}^{}_{\mathrm{vac}} + {\mathcal
H}^{}_{\mathrm{SI}} + {\mathcal
H}^{}_{\mathrm{NSI}} \ ,
\eea 
where ${\mathcal H}^{}_{\mathrm{vac}} $,   ${\mathcal H}^{}_{\mathrm{SI}}$ and ${\mathcal H}^{}_{\mathrm{NSI}}$  represent the Hamiltonian in vacuum and in presence of 
SI  and NSI, respectively.  Thus, 

\bea
 \label{hexpand} 
 {\mathcal
H}^{}_{\mathrm{}} &=& 
\dfrac{1}{2 E} \left\{ {{\mathcal U} \left(
\begin{array}{ccc}
0   &  &  \\  &  \Delta m^2_{21} &   \\ 
 &  & \Delta m^2_{31} \\
\end{array} 
\right) {\mathcal U}^\dagger }_{  } + 
{ {A (x)}   \left(
\begin{array}{ccc}
1  &    &  \\ & 0 &  
 \\  & &0 \\
\end{array} 
\right)}_{   } + 
 {{A (x)}   \left(
\begin{array}{ccc} \varepsilon_{ee}  & \varepsilon_{e \mu}  & 
\varepsilon_{e \tau}  \\ {\varepsilon_{e\mu} }^ \star & 
\varepsilon_{\mu \mu} &   \varepsilon_{\mu \tau} \\ 
{\varepsilon_{e \tau}}^\star & {\varepsilon_{\mu \tau}}^\star 
& \varepsilon_{\tau \tau}\\
\end{array} 
\right)}_{ } 
\right\}  \ , \nonumber\\
 \eea 
where $A (x)=2 E \sqrt{2} G_F n_e (x)$ is the standard CC potential due to
the coherent forward scattering of neutrinos. The three flavour neutrino mixing matrix ${\mathcal
  U}$ $\equiv {\cal U}_{23} \, {\cal W}_{13} \,{\cal U}_{12}$ with
  ${\cal W}_{13} = {\cal U}_\delta~ {\cal U}_{13}~ {\cal
    U}_\delta^\dagger$ and ${\cal U}_\delta = {\mathrm{diag}}
  \{1,1,\exp{(i \delta)}\}$] is characterized by three angles and a
single (Dirac) phase and, in the standard  Pontecorvo-Maki-Nakagawa-Sakata (PMNS) parameterisation, we
have
\bea
{\mathcal U}^{} &=& \left(
\begin{array}{ccc}
1   & 0 & 0 \\  0 & c_{23}  & s_{23}   \\ 
 0 & -s_{23} & c_{23} \\
\end{array} 
\right)   
  \left(
\begin{array}{ccc}
c_{13}  &  0 &  s_{13} e^{- i \delta}\\ 0 & 1   &  0 \\ 
-s_{13} e^{i \delta} & 0 & c_{13} \\
\end{array} 
\right)  \left(
\begin{array}{ccc}
c_{12}  & s_{12} & 0 \\ 
-s_{12} & c_{12} &  0 \\ 0 &  0 & 1  \\ 
\end{array} 
\right)  \ ,
 \eea 
where $s_{ij}=\sin {\theta_{ij}}, c_{ij}=\cos \theta_{ij}$.   It should be noted that  two Majorana phases do not play any role in neutrino
oscillations, and hence not considered.  %

In order to elucidate the role of different NSI terms in a particular oscillation channel,  we can obtain  approximate analytic expressions for  oscillation probabilities  corresponding to various channels
 using techniques of perturbation theory.  The analytic computation of probability expressions in presence of NSI
has been carried out for different experimental settings~\cite{Kopp:2007ne,Kikuchi:2008vq,Asano:2011nj}.  Let us define the following ratios for the sake of convenience, 
\begin{equation}
\lambda \equiv \frac{\Delta m^2_{31}}{2 E}  \quad \quad ; \quad \quad
r_{\lambda} \equiv \frac{\Delta m^2_{21}}{\Delta m^2_{31}} \quad \quad ; 
\quad \quad r_{A} \equiv \frac{A (x)}{\Delta m^2_{31}} \ .
\label{dimless}
\end{equation}
The expressions given below are valid for atmospheric and long
baseline neutrinos where $\lambda L \simeq {\cal O} (1)$ holds and $r_A L
\sim {\cal O} (1)$ for a large range of 
the $E$ and $L$ values.  
The   $\numu \to \nue$ oscillation probability is given by 
\begin{eqnarray}
 P_{\mu e}^{{{NSI} }} &\simeq& 
 4 s_{13}^2 s_{23}^2 \, \left[\dfrac{ \sin^2 {(1-r_A )\lambda L/2}{}}{ (1-r_A)^2} \right] \nonumber\\
 && + \, 8  s_{13} s_{23} c_{23}
( |\varepsilon_{e \mu}| c_{23} c_{\chi} - |\varepsilon_{e \tau}| s_{23} c_{\omega} ) \, r_A \,\left[\dfrac{\sin {r_A \lambda L/2}}{r_A} ~
  \dfrac{ \sin {(1 - r_A) \lambda L/2}}{(1-r_A)} ~\cos \frac{ \lambda L}{2}
 \right] \nonumber\\
  &&
+ \, 8  s_{13} s_{23} c_{23}
( |\varepsilon_{e \mu}| c_{23} s_{\chi} - |\varepsilon_{e \tau}| s_{23} s_{\omega} )r_A \,\left[\dfrac{\sin {r_A \lambda L/2}}{r_A} ~
  \dfrac{\sin {(1 - r_A) \lambda L/2}}{(1-r_A) } ~\sin \frac{ \lambda L}{2}
 \right]\nonumber\\
 && + \, 8  s_{13} s_{23}^2 
( |\varepsilon_{e \mu}| s_{23} c_{\chi} + |\varepsilon_{e \tau}| c_{23} c_{\omega} )r_A \,\left[
 \dfrac{\sin^2 {(1 - r_A) \lambda L/2} }{(1-r_A)^2} 
 \right] \ ,
 \label{pem}
 \end{eqnarray}
where we have used $\tilde s_{13} \equiv \sin \tilde \theta_{13} =
s_{13}/(1-r_A)$ to the leading order in $s_{13}$, and
$\chi=\phi_{e\mu}+ \delta$, $\omega = \phi_{e\tau}+\delta$. Only the
parameters $\varepsilon_{e \mu} $ and $ \varepsilon_{e\tau}$ enter in the
leading order expression~\cite{Kopp:2007ne,Kikuchi:2008vq, Asano:2011nj}, {as} terms such as
$r_\lambda \varepsilon_{\alpha \beta}$ have been neglected.
 
The muon neutrino  survival probability ($\numu \to \numu$) 
is given by
\begin{eqnarray}
\label{pmm}
 P_{\mu\mu}^{{{NSI}}}  &\simeq& 1 - s^2_{2\times {23}} \left[ \sin^2 \frac{\lambda L}{2} \right] \nonumber\\
 && - ~
 |\varepsilon_{\mu\tau}| \cos \phi_{\mu\tau} s_{2 \times {23}} \left[ s^2 _{2 \times {23}} (r_A \lambda L) \sin {{\lambda L}{}} + 4  c^2_{2 \times {23}}  r_A \sin^2 \frac{ \lambda L}{2} 
 \right]\nonumber\\
 && +~ (|\varepsilon _{\mu\mu}| - |\varepsilon _{\tau\tau}|) s^2_{2 \times {23}} c_{2 \times {23}}\left[  \dfrac{r_A \lambda L}{2} \sin  {\lambda L}{} - 2  r_A \sin^2 \frac{\lambda L}{2} \right] \ , 
 \end{eqnarray}
where $s_{2 \times {23}} \equiv \sin 2 \theta_{23} $ and $c_{2 \times
  {23}} \equiv \cos 2 \theta_{23} $. Note that the NSI parameters involving the electron sector do not enter this channel and the survival probability depends only on the three
parameters $\varepsilon_{\mu\mu}, \varepsilon_{\mu\tau},
\varepsilon_{\tau\tau}$.  

\subsection{Leggett-Garg Inequalities}
\label{subsec:lgi}

In order to write down an expression for LGI, we require correlation functions $C_{ij} = \langle \hat Q (t_i) \hat Q (t_j) \rangle$ of a dichotomic observable, $\hat Q (t)$ (with realizations $\pm 1$) at distinct measurement times $t_i$ and $t_j$. Here, $\langle \ldots \rangle$ implies averaging over many trials. As stated in~\cite{Leggett:1985zz}, the actual derivation of LGIs relied on the assumption that measurements of $\hat Q$ at different times $t_i$ are carried out in non-invasive manner.  
However, one may relax this and assume stationarity in which case $C_{ij}$ depend only on the time difference $\tau = t_j - t_i$~\cite{emary2013leggett}.  
 In terms of the joint probabilities,  
the two-time correlation function can be expressed as 
\bea
C_{ij} &=& \sum _{\hat Q_i \hat Q_j = \pm1} \hat Q_i \hat Q_j \mathbb{P}_{\hat Q_i \hat Q_j} (t_i,t_j)
\eea
{where  $\mathbb{P}_{\hat Q_i \hat Q_j} (t_i,t_j)$ is the joint probability of obtaining the results $\hat Q_i$ and $\hat Q_j$ from successive measurements at times $t_i$ and $t_j$, respectively. 
For  $n$-time measurement, we can define the parameter $K_n$ as 
\bea
 K_n &\equiv& \sum _{i=1} ^{n-1} C_ {i,i+1}  - C_{1,n}
\eea
Note that $K_n$ quantifies the extent of violation of LGI. 

The simplest LGIs  are for three-time and four-time measurements ($K_3$ and $K_4$),  which can be expressed as~\cite{emary2013leggett}
\bea
-3 \leq K_{3} \leq 1 \quad {\textrm{where}} \quad K_3 &=& C_ {12} + C_{23} - C_{13} \  \nonumber \\
-2 \leq K_{4} \leq 2 \quad {\textrm{where}} \quad K_4 &=& C_{12} + C_{23} + C_{34} - C_{14}  
\label{classical}
\eea
The generalization  for the case of $n$-time measurements leads to 
\begin{eqnarray}
-n \leq K_{n} \leq n - 2 \quad  {\textrm{}} \quad {\textrm{for odd values of }}\,  n\, (\ge 3) 
\nonumber\\
-(n - 2) \leq K_{n} \leq n - 2 \quad  {\textrm{}} \quad {\textrm{for even values of }}\,  n\, (\ge 4) 
\end{eqnarray}
 Thus, for $n$-time measurements with $n \ge 3$, $K_n \le n-2$ for any value of $n$. 
  These inequalities have been tested in many experiments and are found to be violated~\cite{emary2013leggett}.
The maximum quantum-mechanical value of the LGI correlator  with the standard measurement protocol is 
 \bea
 K_{3}^{max} =\dfrac{3}{2} \quad {\textrm{and}}  \quad K_{4}^{max} =2 \sqrt{2}      \label{eq:bounds}
\eea
These bounds (referred to as Luder's bound or temporal Tsirelson's bounds~\cite{Cirelson:1980ry,PhysRevLett.111.020403,PhysRevLett.113.050401,Rohrlich:1995cf}) are analogous to Tsirelson bounds in the context of spatially separated observations.

Let us try to explain the sense of these bounds for the lowest order LGI parameter, $K_3$. First of all, $-3 \leq K_3 \leq 3$ is the algebraic bound since $C_{ij}$ can take values $\pm 1$.  The bounds in Eq.~\ref{classical}, \ie, $-3 \leq K_3 \leq 1$ imply that (a) the underlying dynamics classical (\ie, macroscopic realism and NIM hold), or  (b)  even if the dynamics is quantum, it does not violate macroscopic realism and NIM which are expected to hold for classical dynamics. So, in this sense, this  represents the classical bound.  If the two conditions (\ie, macroscopic realism and NIM) are violated then the dynamics is quantum and  we expect  $1 < K_{3}  \leq 1.5 $  and the upper bound in this case  is called the Luder's bound or temporal Tsirelson's bound as stated above. 
 Hence, the LGI parameter values lying outside the classical limits (\ie, if $1 \leq K_{3}  \leq 1.5$) are indicative of the quantumness.
\subsection{Neutrino oscillations and LGI violation}
\label{subsec:lginu}

In sharp contrast to the electronic or photonic systems,   neutrinos exhibit coherence over 
astrophysical length scales which offers us a unique setting to
  test LGI in neutrino oscillations. Tests of LGI have been carried out on neutrinos~\cite{Formaggio:2016cuh,Fu:2017hky}. 
  Let us describe the formalism to compute the 
  joint probabilities for two- and three- flavour neutrino oscillations.   
  
 The initial state is taken to be $|\nu_e \rangle$. {\hlpm{Electron neutrinos could be 
 produced in the Beta beam set-up~\cite{betabeam}}}.     The dichotomic observable,
   $\hat Q$ assumes value $+1$ if neutrino is electron flavoured, $|\nu_e \rangle$. $\hat Q$ assumes value 
   $-1$ if neutrino is muon flavoured, $|\nu_\mu \rangle$ (or tau flavoured, $|\nu_\tau \rangle$ in the three flavour case).
   
We first briefly review the simplest case of two flavour neutrino oscillations in vacuum.  We can express $C_{12}$  as 
\begin{eqnarray}
C_{12} &=& \mathbb{P}_{\nu_{e} \nu_{e}}(t_1,t_2)-\mathbb{P}_{\nu_{e} \nu_{\mu}}(t_1,t_2)-\mathbb{P}_{\nu_{\mu} \nu_{e}}(t_1,t_2)+\mathbb{P}_{\nu_{\mu} \nu_{\mu}}(t_1,t_2)   \nonumber
\end{eqnarray}
 {where {{$\mathbb{P}_{\nu_{\alpha} \nu_{\beta}} (t_1,t_2) = P_{\nu_
e \to \nu_\alpha} (t_1) P_{\nu_\alpha \to \nu_\beta} (t_2)$}} is the 
joint probability of obtaining  neutrino in state $|\nu_\alpha\rangle$ at time 
$t_{1}$ and in state $|\nu_\beta \rangle $ at time $t_{2}$.  
Using the two flavour probability expressions in vacuum~\cite{giunti}, it can be shown that $C_{12}$ takes the form~\cite{Gangopadhyay:2013aha}  
\begin{eqnarray}
C_{12} &=& 1 - 2 \sin^2 2\theta \sin^2 \left(\dfrac{\lambda }{2} \Delta L \right)  
 \label{c12_vac} 
\end{eqnarray}
where $  \Delta L \equiv L_2 - L_1  $  (with $L_1$ and $L_2$ being the distance from the source at which measurements take place)  has been used in place of $\tau \equiv t_2-t_1$ (in the ultra-relativisitic limit, we use the substitution $t \to$ $L$).  Thus, we note that in  the two flavour case,  $C_{12}$ depends only on $\Delta L \equiv  L_2 - L_1 $ and not on individual $L_1$ or $L_2$,  thereby respecting  the stationarity condition naturally. We can compute the other $C_{ij}$'s in a similar way and  evaluate $K_4$.
If we take all separations to be equal, i.e., $L_2-L_1 = L_3 - L_2 = L_4 - L_3 \equiv \Delta L$, {\hlpm{we have $C_{12} = C_{23} = C_{34}$. Under this assumption,}}
$K_4$ is given by 
\begin{eqnarray}
K_4 &=& 2 - 2 \sin^2 2\theta \left[ 3 \sin ^2 \left(\dfrac{\lambda}{2} \Delta L \right)   -  \sin ^2 \left(\dfrac{3 \lambda  }{2} \Delta L \right)   \right]
 \label{k4_vac} 
\end{eqnarray}
Interestingly, it can be noted from Eq.~\ref{k4_vac}, that $K_4$ is an oscillatory function of $\Delta L/E$. We will demonstrate this in Sec.~\ref{sec:res}.

The earth matter effects (both SI and NSI) modify the effective angle, $\theta \to \tilde \theta$ and mass-squared difference, $\Delta m^2 \to \tilde {\Delta m^2} $, but the form of probabilities remain~\cite{giunti} and therefore $C_{12}$ (and $K_4$) in matter will also have the same form as in vacuum (see Eqs.~\ref{c12_vac} and  \ref{k4_vac}) and consequently, stationarity will remain preserved even in presence of matter.

In the context of three flavour neutrino oscillations in matter,    $C_{12}$ can be written down in terms of the nine joint probabilities~\cite{Gangopadhyay:2017nsn},
\begin{eqnarray}
C_{12}
&=& \mathbb{P}_{\nu_{e} \nu_{e}}(L_1,L_2)-\mathbb{P}_{\nu_{e} \nu_{\mu}}(L_1,L_2)-\mathbb{P}_{\nu_{e} \nu_{\tau}}(L_1,L_2)
- \mathbb{P}_{\nu_{\mu} \nu_{e}}(L_1,L_2)+\mathbb{P}_{\nu_{\mu} \nu_{\mu}}(L_1,L_2) 
\nonumber\\
&& +\, \mathbb{P}_{\nu_{\mu} \nu_{\tau}}(L_1,L_2) - \mathbb{P}_{\nu_{\tau} \nu_{e}}(L_1,L_2)+\mathbb{P}_{\nu_{\tau} \nu_{\mu}}(L_1,L_2)+\mathbb{P}_{\nu_{\tau} \nu_{\tau}}(L_1,L_2)
\label{lgi}
\end{eqnarray}
where $\mathbb{P}_{\nu_{\alpha} \nu_{\beta}} (L_1,L_2) = P_{\nu_
e \to \nu_\alpha} (L_1) P_{\nu_\alpha \to \nu_\beta} (L_2)$ is the 
joint probability of obtaining a neutrino in state $|\nu_\alpha\rangle$ at  
$L_{1}$ and in state $|\nu_\beta \rangle $ at  $L_{2}$. 
Using the approximate expressions of oscillation probabilities in matter with SI, it is straightforward but somewhat tedious to 
compute $C_{12}$ and other correlation functions. 
 $C_{12}$  is given by
\begin{eqnarray}
 C_{12} && = 
  \Bigg[1-r^{2}_{\lambda}\sin ^{2}2\theta_{12}\frac{\sin ^{2}
\big(\frac{r_{A} \lambda L_{1}}{2}\big)}{\Big(r_{A}
\Big)^{2}}-4s^{2}_{13}\frac{\sin ^{2}(r_{A}-1)\frac{\lambda L_{1}}{2}}
{(r_{A}-1)^{2}}\Bigg]
\Bigg[1-2r_{\lambda} 
^{2}\sin ^{2}2\theta_{12}\frac{\sin ^{2}\big(\frac{r_{A} \lambda \Delta L}{2}
\big)}{r_{A}^{2}}
 \nonumber\\
 &&-8s^{2}_{13}\frac{\sin ^{2}(r_{A}-1)\frac{\lambda \Delta L}{2}}
 {(r_{A}-1)^{2}}\Bigg]-\Bigg[r_{\lambda} 
 ^{2}\sin ^{2}2\theta_{12}c^{2}_{23}\frac{\sin ^{2}\big(\frac{VL_{1}}
 {2}\big)}{r_{A}^{2}}+4s^{2}_{13}
 s^{2}_{23}\frac{\sin ^{2}(r_{A}-1)\frac{\lambda L_{1}}{2}}
 {(r_{A}-1)^{2}}  +2r_{\lambda} s_{13} \nonumber\\  &&\sin 2\theta_{12}\sin 2\theta_{23} \cos \Big(\frac{\lambda L_{1}}{2} -\delta \Big)\frac{\sin \big(\frac{r_{A} \lambda L_{1}}{2}\big)}{r_{A}}\frac{\sin(r_{A}-1)\frac{\lambda L_{1}}{2}}{(r_{A}-1)}\Bigg]\Bigg[2r^{2}_{\lambda}\sin ^{2}2\theta_{12}c^{2}_{23}\frac{\sin ^{2}\big(\frac{r_{A} \lambda \Delta L}{2}\big)}{r_{A}^{2}}\nonumber\\
&&+8s^{2}_{13}s^{2}_{23}\frac{\sin ^{2}(r_{A}-1)\frac{\lambda \Delta L}{2}}{(r_{A}-1)^{2}}+4r_{\lambda} s_{13}\sin 2\theta_{12}\sin 2\theta_{23} \frac{\sin \big(\frac{r_{A} \lambda \Delta L}{2}\big)}{r_{A}} \frac{\sin (r_{A}-1)\frac{\lambda \Delta L}{2}}{(r_{A}-1)}\nonumber\\
&&\times \Big\{\cos \Big(\frac{\lambda \Delta L}{2} -\delta \Big)-\sin \delta _{}\sin \Big(\frac{\lambda \Delta L}{2}\Big)\Big\}-1\Bigg]-\Bigg[r^{2}_{\lambda}\sin ^{2}2\theta_{12}s^{2}_{23}\frac{\sin ^{2}\big(\frac{r_{A} \lambda L_{1}}{2}\big)}{r_{A}^{2}}+4s^{2}_{13}c^{2}_{23}\nonumber\\
&&\times \frac{\sin ^{2}(r_{A}-1)\frac{\lambda L_{1}}{2}}{(r_{A}-1)^{2}}-2r_{\lambda} s_{13}\sin 2\theta_{12}\sin 2\theta_{23} \cos \Big(\frac{\lambda L_{1}}{2} -\delta \Big)\frac{\sin \big(\frac{r_{A} \lambda L_{1}}{2}\big)}{r_{A}}\frac{\sin \Big\{(r_{A}-1)\frac{\lambda L_{1}}{2}\Big\}}{(r_{A}-1)}\Bigg]\nonumber\\
&&\Bigg[2r^{2}_{\lambda}\sin ^{2}2\theta_{12}s^{2}_{23}\frac{\sin ^{2}\big(\frac{r_{A} \lambda \Delta L}{2}\big)}{r_{A}^{2}}+8s^{2}_{13}c^{2}_{23}\frac{\sin ^{2}\Big\{(r_{A}-1)\frac{\lambda \Delta L}{2}\Big\}}{(r_{A}-1)^{2}}-4r_{\lambda} s_{13}\sin 2\theta_{12}\sin 2\theta_{23}\nonumber\\
&&\frac{\sin \big(\frac{r_{A} \lambda \Delta L}{2}\big)}{r_{A}} \frac{\sin (r_{A}-1)\frac{\lambda \Delta L}{2}}{(r_{A}-1)} \Big\{\cos \Big(\frac{\lambda \Delta L}{2} -\delta \Big)-\sin \delta \sin \Big(\frac{\lambda \Delta L}{2}\Big)\Big\}-1\Bigg]
\label{c12_mat}
\end{eqnarray}
 We note that $C_{12}$  has explicit dependence on individual baseline, $L_{1}$ as well as the spatial separation, $\Delta L = L_2- L_1$.   Thus, stationarity condition is violated in the three flavour case when standard matter interactions are taken into consideration. We can compute the other $C_{ij}$'s in a similar way.      Using the approximate expression of probabilities in presence of NSI, writing down the  expression for $C_{12}$ in a compact form is a difficult task as there are large number of
      additional  parameters. Moreover, it would  not lead to any further insight on the nature of correlations or violation of the stationarity condition. The relative dependence on the NSI parameters can be gleaned from the  expressions for oscillation probabilities given in Sec.~\ref{sec:nsi}.

\section{Results}
\label{sec:res}

\begin{figure}[t!]
\centering
\includegraphics[width=.9\textwidth]{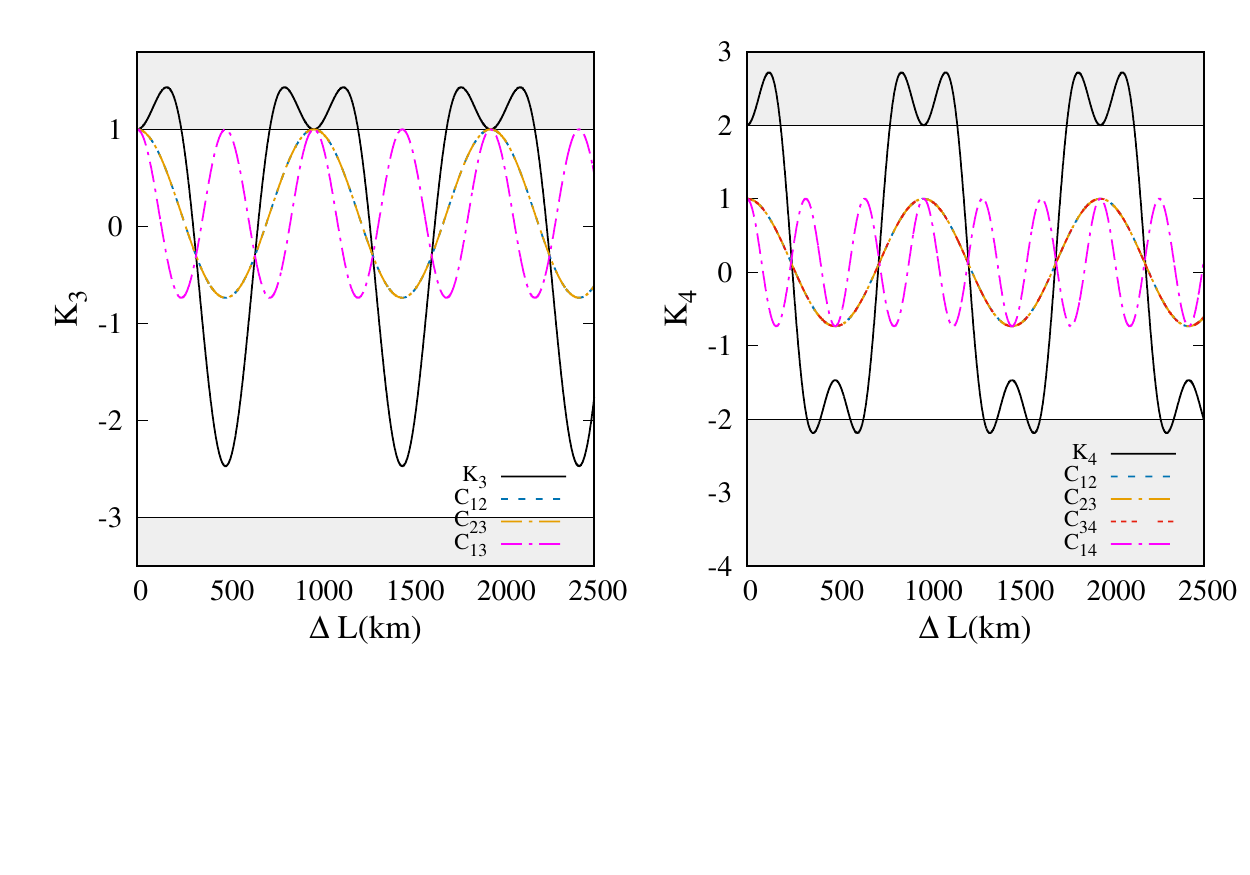}
\vskip -.6in
\caption{$K_3$ and $K_4$  plotted as a function of $\Delta L$ for the two flavour neutrino oscillations in vacuum.   The contribution of various $C_{ij}$'s is depicted in the two panels.
  The grey shaded regions imply violation of LGI in this and the following figures. %
  }
\label{fig_a}
\end{figure}
 
In this section, we use the latest best-fit values of the oscillation parameters given in Table~\ref{tab}. The neutrino mass ordering can be normal (NO) or inverted (IO). 
We assume NO for all the plots, unless stated otherwise. The LGI quantities $K_3 $ and $K_4$ are then obtained using the joint probabilities. 
   We take the initial neutrino flavour to be $| \nu_e \rangle$. Note that  the energy is held fixed  at  $E=1$ GeV in all the plots, unless stated otherwise.

  To begin with, we first plot $K_3$ and $K_4$ as a function of $\Delta L$ for the case of two flavour neutrinos in Fig.~\ref{fig_a}.  If we examine Eq.~\ref{k4_vac} carefully, we expect to get an oscillatory function of $\Delta L$ (for fixed $E$).  
 The exact numerical results agree  perfectly well with this observation.   Clearly, there is an interplay of various $C_{ij}$'s which leads to an overall shape of  $K_3$ and
   $K_4$ curves as a function of $\Delta L$  (see Fig.~\ref{fig_a}).  {\hlpm{From Fig.~\ref{fig_a}, we note that    $C_{12}$ (taken to be equal to $C_{23} $) dictates the overall frequency of $K_3$ and $C_{13}$ leads to fine features. Likewise,  $C_{12}$ (taken to be equal to $C_{23} $ and $C_{34}$)  dictates the overall frequency of the $K_4$ and $C_{14}$ leads to fine features. }}
The joint probability terms appearing in  $C_{13}$ and $C_{14} $ are responsible for the fine features of the overall curve (in black) for $K_3$ and $K_4$ respectively. 
   We take $L_{4} - L_{3} = L_{3} - L_{2} = L_{2} - L_{1} \equiv \Delta L$ (see Ref.~\cite{Gangopadhyay:2013aha}).
 The largest violation of LGI  is observed at the following values of $\Delta L$ :
 \begin{description}
\item $K_{3}^{m, 2fl} \simeq 1.433$ at $\Delta L \simeq 160$ km, 
\ldots $2090$ km
\item $K_{4} ^{m, 2fl} \simeq 2.718 $ at $\Delta L \simeq 120$ km,  \ldots $2050$ km
\end{description}
It should be noted that in the two flavour case,  the values of  $K_{3}^{m,2fl}$ and $K_{4}^{m,2fl}$ are just below the respective maximal attainable bounds as  given in Eq.~\ref{eq:bounds}.

In order to assess the role of NSI  on the LGI for  three flavour neutrino oscillations, we compute the oscillation probabilities  numerically using the General Long Baseline Experiment Simulator (GLoBES) and associated implementation of NSI~\cite{Huber:2004ka, Huber:2007ji}.  The matter density has been taken to be $3$ g/cc. 

\begin{figure}[t!]
\centering
\includegraphics[width=.99\textwidth] {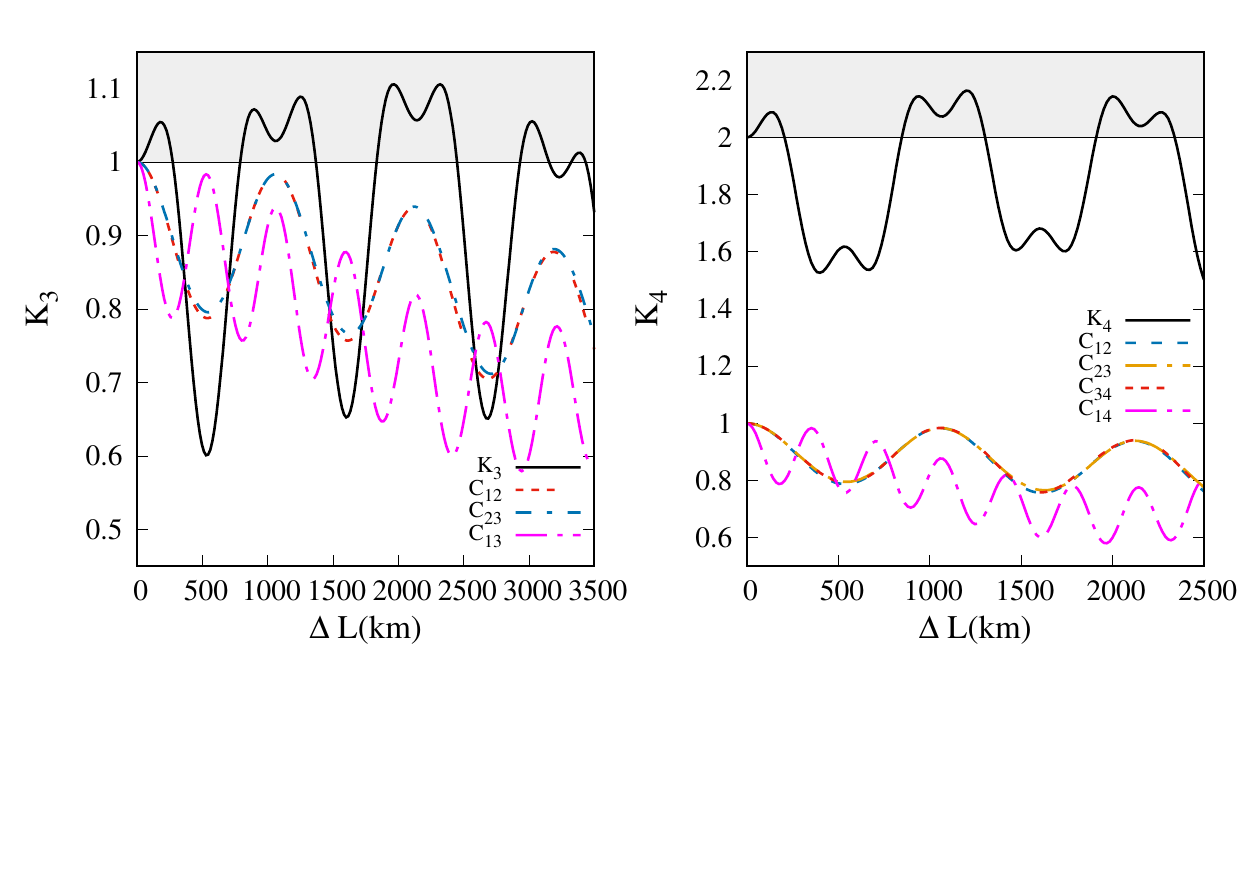}
\vskip -0.8in
\caption{$K_3$ and $K_4$  plotted as a function of $\Delta L$ for the three flavour neutrino oscillations in matter with SI. The contribution of various $C_{ij}$'s is depicted in the two panels.}
\label{fig_b}
\end{figure}

\begin{figure}[t!]
\centering
\includegraphics[width=.6\textwidth] {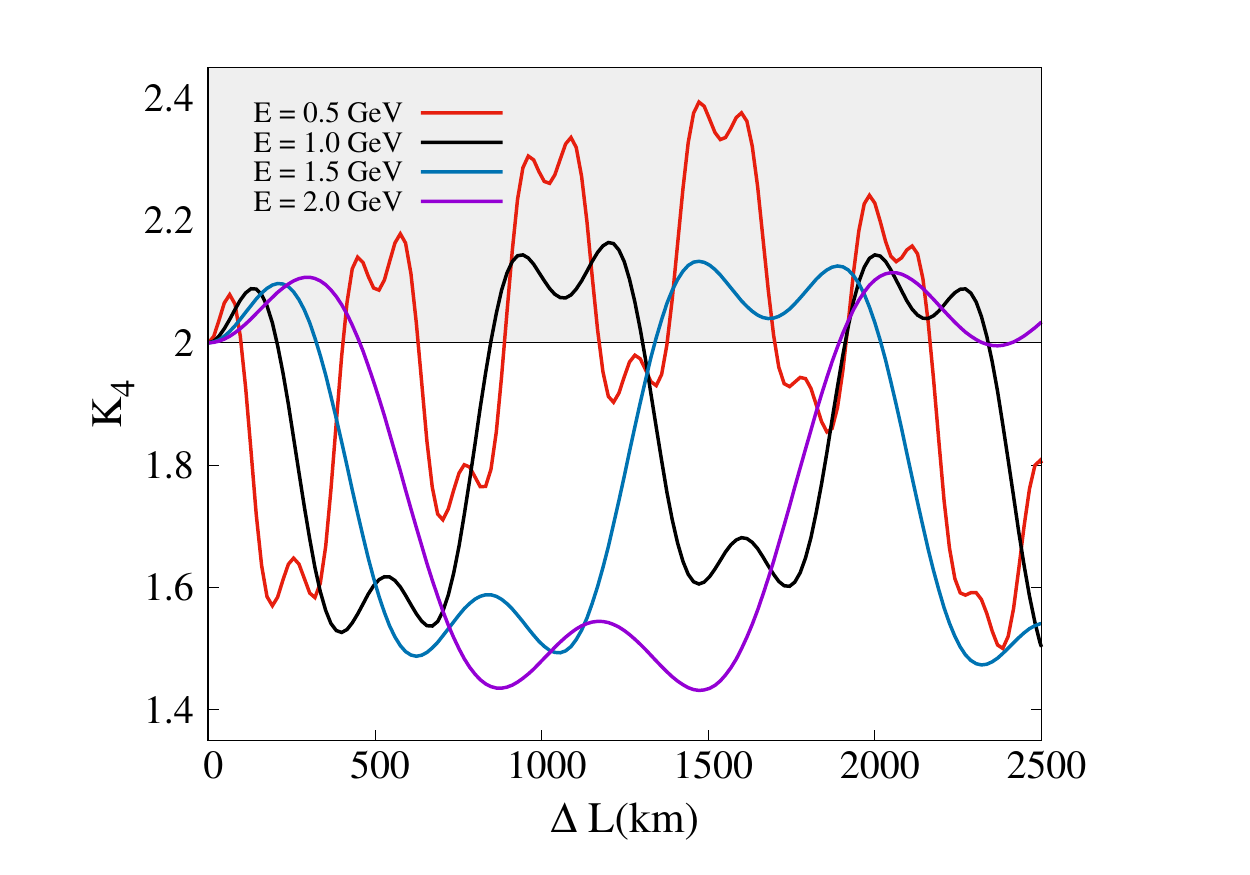}
\caption{$K_4$  plotted as a function of $\Delta L$ for the three flavour neutrino oscillations in matter (SI case) for different values of $E$ taken to be $0.5, 1.0, 1.5$ and $2$ GeV.} 
\label{fig_new}
\end{figure}

 For the three flavour oscillations in matter with SI, the quantities $K_3 $ and $K_4$ are plotted  as function of $\Delta L$ in Fig.~\ref{fig_b}. 
Going from two- to three- flavour oscillations as shown in Fig.~\ref{fig_b}, we first note that the nice oscillatory feature is lost. This can be understood from the fact that stationarity condition is no longer maintained in this case.  Also, in contrast to the two flavour case, the range of allowed values of $K_3 $ and $K_4$  are modified.  
As mentioned earlier,   the shape of the curves for $K_3$ and $K_4$   in case of SI  are  dictated by 
 the interplay of the  $C_{ij}$'s.   Here again $L_{4} - L_{3} = L_{3} - L_{2} = L_{2} - L_{1} \equiv \Delta L$ and  $L_{1}$ is chosen to be 140.15 km which maximises the LGI parameters, $K_3$ and $K_4$~\cite{Gangopadhyay:2017nsn}.    The maximum violation of LGI in the standard case is found at the following values of $\Delta L$ :
\begin{description}
\item $K_{3}^{m} \simeq 1.106$ at $\Delta L \simeq 1970$ km \& $2320$ km
\item $K_{4} ^{m} \simeq 2.163 $ at $\Delta L \simeq 1200 $ km
\end{description}
Thus, as we go from two to three flavour case, the maximum values of LGI parameters, $K_{3}^{m}$ and $K_{4}^{m}$ are much smaller than the corresponding values ($K_{3}^{m,2fl}$ and $K_{4}^{m,2fl}$) obtained in the two flavour case.

\begin{figure}[t!]
\centering
\includegraphics[width=.99\textwidth] {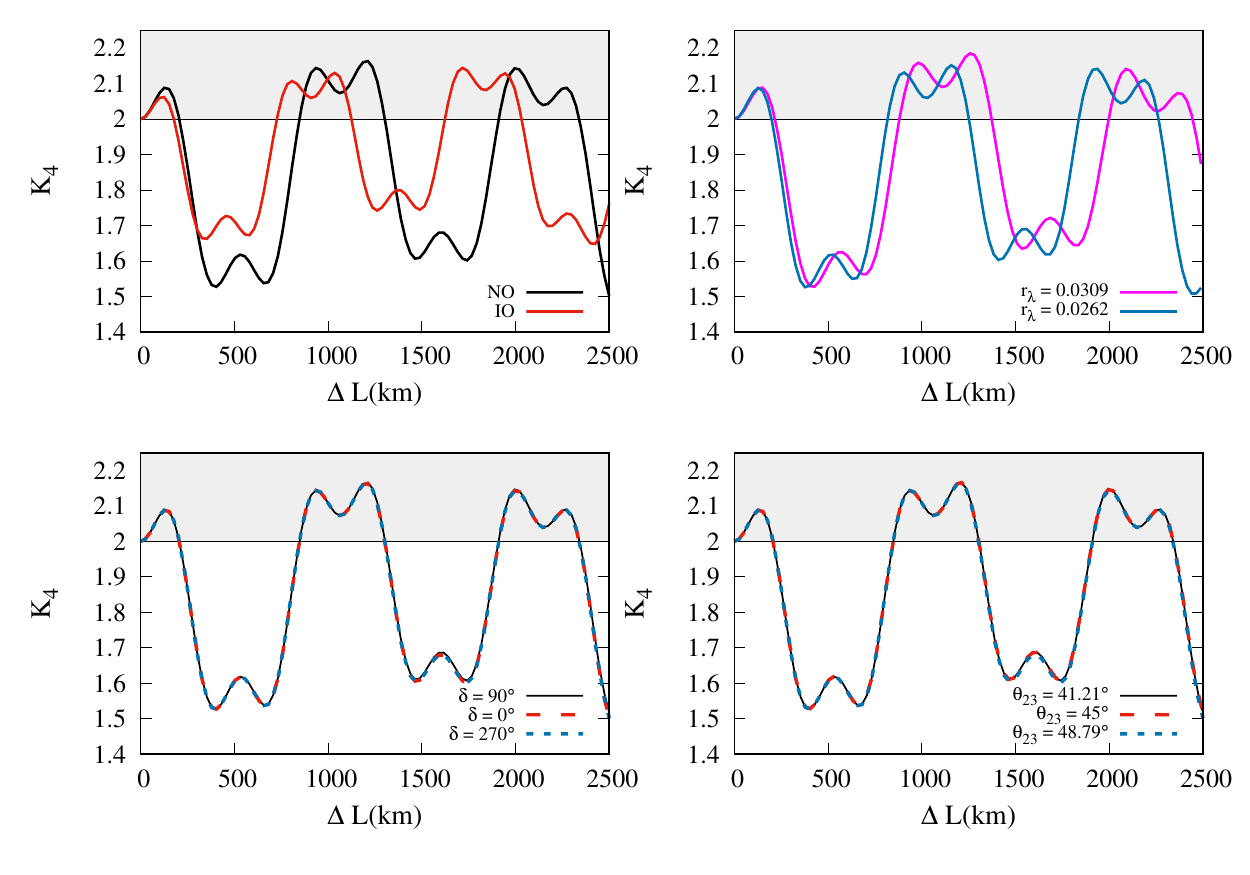}
\label{fig_c}\caption{The dependence of $K_4$ on the  current unknowns in neutrino oscillation physics - the ordering of neutrino masses (sign of $\Delta m^2_{31}$), the 
dependence on value of mass ordering parameter, $r_\lambda$ in the allowed range (for NO),  the value of the CP violating phase, $\delta$ and  the octant of $\theta_{23}$   as a function of $\Delta L$ for three flavour neutrino oscillations in presence of SI. 
  }
\end{figure}

In Fig.~\ref{fig_new}, we show the dependence of $K_4$  on the energy ($E = 0.5, 1.0, 1.5, 2.0$ GeV).  {\hlpm{Note that the $E$-dependence arises via factors such as $\lambda\propto 1/E$ and $r_A \propto E$ appearing in  Eq.~\ref{c12_mat}.}} 
 \hlpm{{It can be noted that the value of $K_4$  increases as we go to lower energies. }}
Moreover, there is a shift in the value of $\Delta L$ at which the maximum violation occurs {\hlpm{as the phase factors get modified with change in $E$}}. 
The maximum violation occurs for $E=0.5$ GeV at $\Delta L \simeq 1500$ km.
\begin{description}
\item $K_{4}^{m} \simeq 2.393 $ at $\Delta L \simeq 1470 $ km for $E = 0.5$ GeV
\item $K_{4} ^{m} \simeq 2.163 $ at $\Delta L \simeq 1200 $ km for $E = 1.0$ GeV
\item $K_{4}^{m} \simeq 2.133 $ at $\Delta L \simeq 1470 $ km for $E = 1.5$ GeV
\item $K_{4}^{m} \simeq 2.115 $ at  $\Delta L \simeq 2050 $ km for $E = 2.0$ GeV
\end{description}

Having found the location of maximum violation of LGI  for two- and three- flavour neutrino oscillations, we first examine if the values of $K_4$ has any dependence on the the current unknowns in neutrino oscillation physics in Fig.~\ref{fig_c}. These unknowns are: (i) ordering of neutrino masses (sign of $\Delta m^2_{31}$) as shown in top row (left panel), (ii) the value of the CP violating phase, $\delta$ as shown in bottom row (left panel), and (iii)  the octant of $\theta_{23}$ as shown in bottom row (right panel). %
It is seen that the height of the curve and shift in the location  of maximum violation depends on the ordering of neutrino masses (sign of $\Delta m^2_{31}$).  In general, for IO,  the maximum value that $K_4$  attains is lower and given by $K_4^m \simeq 2.144$ at $\Delta L \simeq 1710$ km. 
There is mild dependence on  value of $\delta$ and $\theta_{23}$. Besides, in the top row (right panel), we also show  the dependence on the value of mass ordering parameter, $r_\lambda$ by varying its value within the allowed range for a given ordering (NO). It is noted that there is some dependence on the value of this parameter as it leads to  change in the location (value of $\Delta L$) of maximum violation  as well as change in value of $K_4^m$.
In order to understand the precise role of antineutrinos for a given ordering (NO) of neutrino masses, we expect that the curve for antineutrinos with NO would be similar   to  neutrinos with IO. This can be understood as follows. The effective combination entering the oscillation probability at the leading order~\cite{giunti} is $r_A$ which is the ratio of $A$ and $\Delta m^2_{31}$ and it does not matter whether we change $A \to - A$ (neutrinos $\to$ antineutrinos) or  the  ordering of neutrino masses  $\Delta m^2_{31} \to - \Delta m^2_{31}$ (NO $\to$ IO).

 \begin{figure}[t!]
\centering 
\includegraphics[width=.6\textwidth] {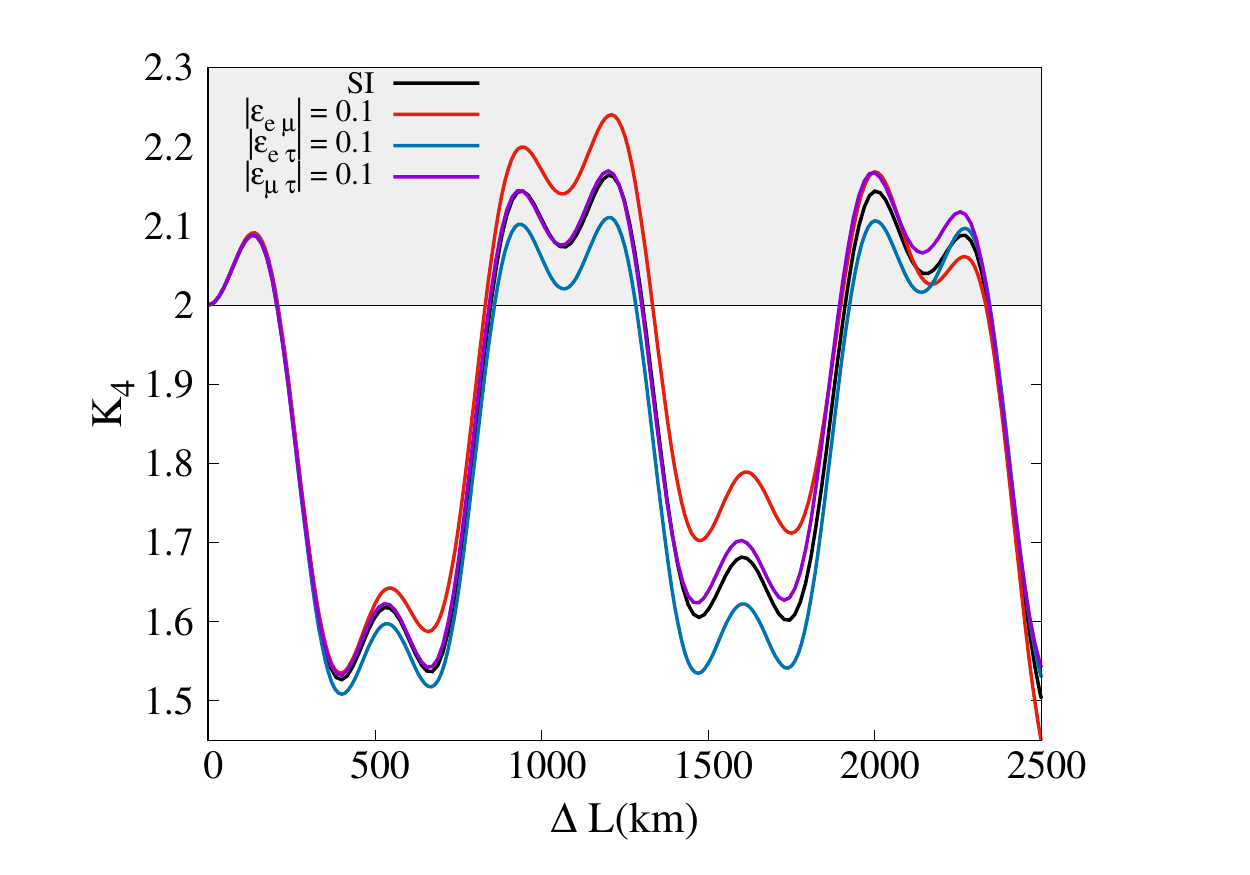}
\caption{$K_4$ is plotted as a  function of $\Delta L$ for three flavour oscillations in presence of NSI.
Here, we take one NSI parameter non-zero at a time. The NSI phases are set to zero. 
}
\label{fig_d}
\end{figure}

Next we analyse the impact of NSI  on the LGI parameter, $K_4$.  We start by considering one NSI parameter non-zero at a time. This simplifies the computation and allows for  a clear understanding of the role played by the specific NSI parameters.  In Fig.~\ref{fig_d},
 we show the impact of $|\varepsilon_{e\mu}|$, $|\varepsilon_{e\tau}|$ and $|\varepsilon_{\mu\tau}|$.  We choose the strength of NSI terms to be $|\varepsilon_{\alpha\beta}|=0.1$ for all the three NSI parameters so that we can compare their impact on similar footing.   As can be noted from Fig.~\ref{fig_d}, the NSI terms,
    $|\varepsilon_{e\mu}|$ and $|\varepsilon_{e\tau}|$ act in the opposite directions which is expected from the analytic expressions of oscillation probabilities in the $\nu_\mu \to \nu_e$ channel given in Sec.~\ref{sec:nsi}~\cite{Masud:2015xva}  and the difference is prominently seen in the grey shaded region that depicts violation of LGI.  While   $|\varepsilon_{e\mu}|$ leads to enhancement in the amount of violation of LGI with respect to the SI case,   $|\varepsilon_{e\tau}|$ leads to suppression.
As noted from Sec.~\ref{sec:nsi},  the parameter $|\varepsilon_{\mu\tau}|$   enters the expression for  $\nu_\mu \to \nu_\mu$ channel. We find that the dependence of $K_4$ on 
     $|\varepsilon_{\mu\tau}|$  is mild and closer to the expectation for the SI case. 
     The NSI phases have been set to zero. This observed enhancement is the key point of this article. The maximum violation of LGI in the NSI case are observed at the following values of $\Delta L$ :
    \begin{description}
    \item $K_{4}^{{m}} \simeq 2.240 $ $(2.331)$ for $|\varepsilon_{e\mu}| = 0.1$ $(0.2)$ at $\Delta L \simeq 1210$ km
\end{description}
It should be noted that  $K_{4}^{m}$ increases with the increase in the absolute value of the NSI term  but still the value is below the maximal attainable bound (see Eq.~\ref{eq:bounds}).
The impact of non-zero NSI phases ($\delta_{e\mu}$ and $\delta_{e\tau}$) while taking one NSI parameter non-zero at a time has been shown in Fig.~\ref{fig_e}. It can be seen that maximum violation in case of non-zero $\varepsilon_{e\mu}$   occurs for $\delta_{e\mu}=0$.
Also, the maximum violation  in case of non-zero $\varepsilon_{e\tau}$   occurs for $\delta_{e\tau}=\pi$.

Motivated by the recent work on explaining the discrepancy between T2K and NoVA using large NSI~\cite{Chatterjee:2020kkm,Denton:2020uda}, we have plotted $K_4$ for the considered values of NSI terms.  
The   impact of  collective NSI terms ($\varepsilon_{e\mu}$ and $\varepsilon_{e\tau}$ along with non-zero phases, $\delta_{e\mu}$ and $\delta_{e\tau}$) is shown in Fig.~\ref{fig_f}. With the goal of maximizing the $K_4$, the phases have been appropriately chosen using the insight obtained above (Fig.~\ref{fig_e}).  The maximum violation of LGI in the NSI case (with $|\varepsilon_{e\mu}  | = 0.2$, $|\varepsilon_{e\tau} |= 0.2$, $\delta_{e\mu} = 0$ and $\delta_{e\tau} = \pi$) is observed at the following value of $\Delta L$ :
\begin{description}
\item $K_{4}^{m} \simeq 2.481$ at $\Delta L \simeq 1200$ km for $\nu$ ($|\varepsilon_{e\mu}|=|\varepsilon_{e\tau}|=0.2$, $\delta_{e\mu}=0$, $\delta_{e\tau}=\pi$) 
\item $K_{4}^{m} \simeq 2.388 $ at $\Delta L \simeq 1000 $ km for $\bar\nu$ ($|\varepsilon_{e\mu}|=|\varepsilon_{e\tau}|=0.2$, $\delta_{e\mu}=\pi$, $\delta_{e\tau}=0$)
\end{description}
By making appropriate choice of NSI parameters, the value of $K_4^m$ can reach upto a maximum of $\sim 2.481$ for $\Delta L \sim 1200$ km. In order to understand the case of antineutrinos for the same ordering,  we know that the sign of the matter potential gets reversed as we go from neutrinos to antineutrinos, \ie, $A \to - A$.  Thus, the choice of NSI phases has to be carefully done, to see effects of enhancement in LGI violation.
Finally, our key results are summarized in Table~\ref{tab:summary}.   It should be noted that we see an enhancement in the value of $K_{4}^{m}$ from $2.163$ (in case of SI) to $2.481$ (for the considered NSI parameters) at around the same value of $\Delta L \sim 1200$ km. This amounts to $\sim 15\%$ relative change. However, as expected, the enhanced value still   stays below the maximal attainable bound of $2 \sqrt{2}$ (see Eq.~\ref{eq:bounds}).

 \begin{figure}[t!]
\centering 
\includegraphics[width=.99\textwidth] {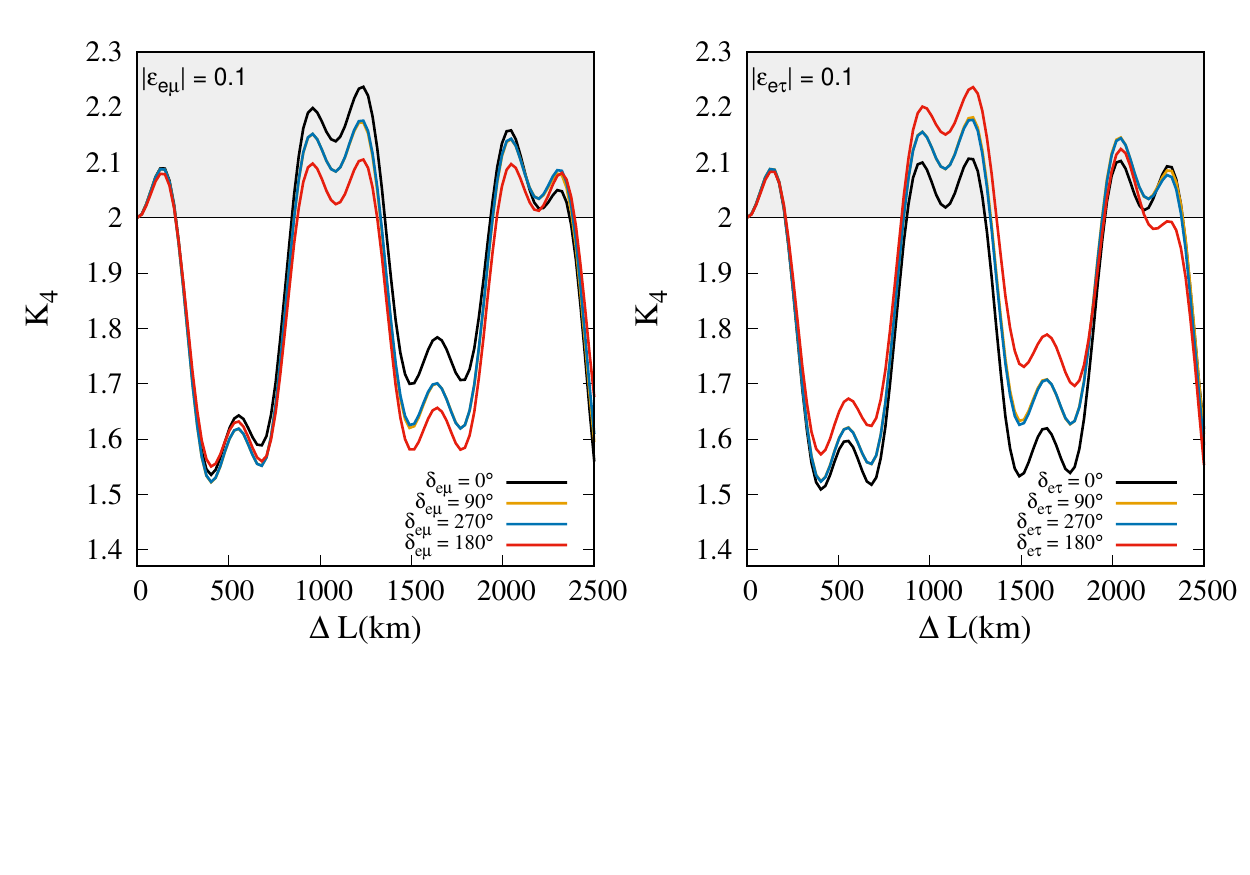}
\vskip -.6in
\caption{$K_4$ is plotted as a  function of $\Delta L$ for three flavour oscillations in presence of NSI. Here, we take one NSI parameter non-zero at a time. The impact of corresponding NSI phases is depicted.}
\label{fig_e}
\end{figure}

\begin{figure}[t!]
\centering
\includegraphics[width=.99\textwidth] {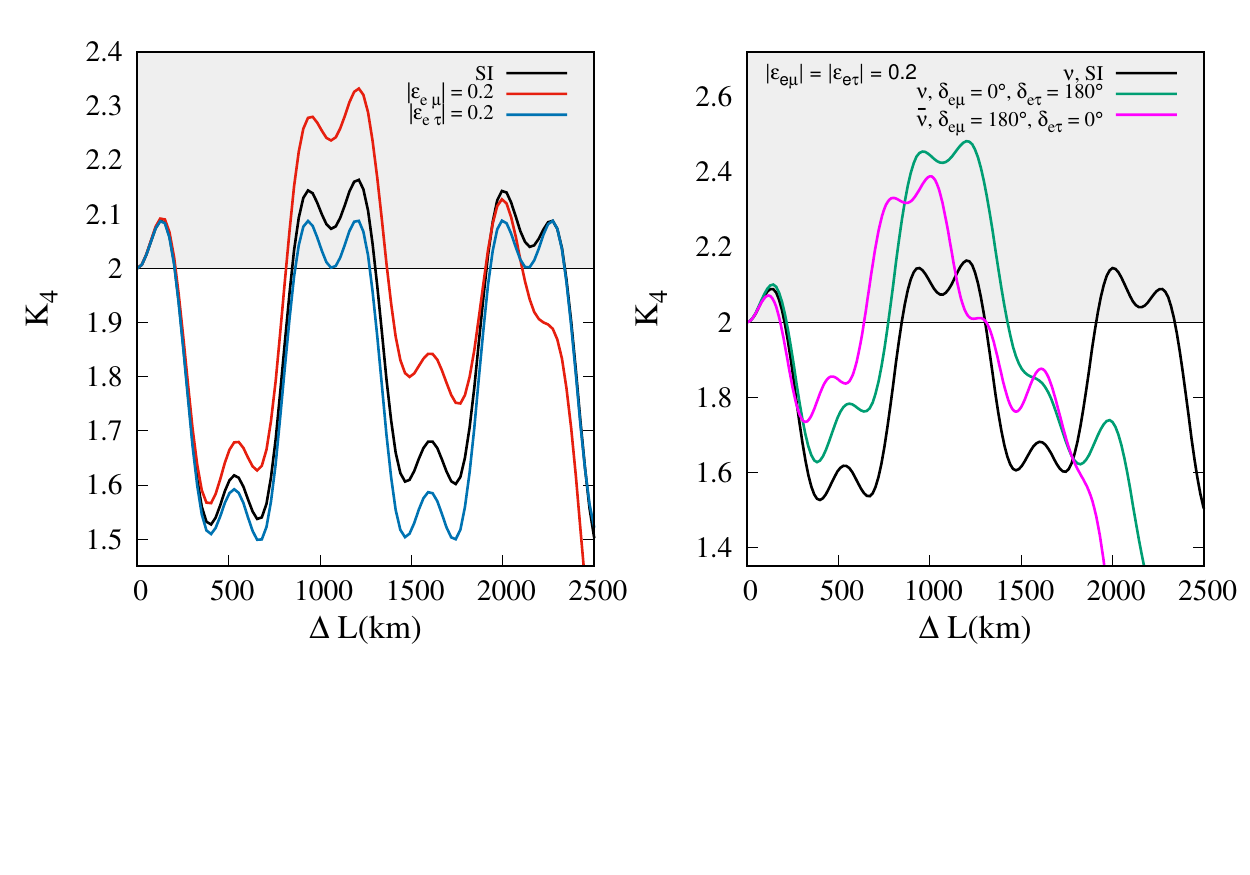}
\vskip -.6in
\caption{$K_4$ is plotted as a  function of $\Delta L$ for three flavour oscillations in presence of NSI. 
Here, we take large values of NSI parameters. The role of individual NSI terms (taken to be real) in order to maximize $K_4$ is shown in the left panel. The role of collective NSI terms along with appropriate phase values in order to maximize $K_4$ is shown in the right panel. 
 }
\label{fig_f}
\end{figure}

 \begin{table}[htb!]
\centering
\begin{tabular}{ l  c  c c c }
\hline
&&& \\
Scenario &  $K_4^m$ & $\Delta L$ (km)  &  Relative change (\%) \\
&& &&\\
\hline
&&&& \\
$\nu$,   SI   & 2.163 & $1200 $ & 0 \\
$\nu$,    NSI ($|\varepsilon_{e\mu}|=0.1$)      &  2.240 & $  1210$  & 3.56     \\
$\nu$,  NSI ($|\varepsilon_{e\mu}|=0.2$)    & 2.331 & $  1210$ & 7.77  \\
$\nu$, NSI ($|\varepsilon_{e\mu}|=|\varepsilon_{e\tau}|=0.2$, $\delta_{e\mu}=0$, $\delta_{e\tau}=\pi$)   & {\bf{2.481}} & $  1200$ & {\bf{14.7}}   \\
$\bar \nu$, NSI ($|\varepsilon_{e\mu}|=|\varepsilon_{e\tau}|=0.2$, $\delta_{e\mu}=\pi$, $\delta_{e\tau}=0$)   & {{2.388}} & $  1000$ &  10.40   \\
&&\\
\hline
\end{tabular}
\caption{Maximum value of $K_4$ and corresponding value of $\Delta L$ for different scenarios. The relative change with respect to SI case (for neutrinos) is also mentioned in the table. NO is assumed. }
\label{tab:summary}
\end{table}

\section{Conclusion and outlook}
\label{sec:con}

That neutrino oscillations is quantum mechanical phenomenon is well-known. Unlike photons, they have extremely large mean free path and exhibit sustained coherence over astrophysical length scales. This provides for a unique opportunity to employ neutrinos as a useful quantum resource. However, in order to do so, one would require a thorough understanding of the aspects such as how neutrinos affect implications of LGI and its violation under different situations - propagation in vacuum as well as in matter with and without standard interactions. In this respect, there are several theoretical studies discussing two~\cite{Gangopadhyay:2013aha,Formaggio:2016cuh,Fu:2017hky} and three~\cite{Gangopadhyay:2017nsn,Naikoo:2017fos,Naikoo:2019eec,Naikoo:2019gme} flavour neutrino oscillations with and without the assumption of stationarity. Additionally, data from two experiments, MINOS and Daya Bay~\cite{Formaggio:2016cuh,Fu:2017hky} have been analysed using a two flavour approach in the context of LGI and a convincing result has been derived from these studies.

Currently, one of the primary goals of the ongoing and future experiments in neutrino oscillation physics are to get a handle on the unknowns - the CP phase ($\delta$), neutrino mass ordering (sign of $\Delta m^2_{31}$) and  the octant of $\theta_{23}$. Moreover, the precision of these experiments would allow for testing the existence of new physics in the form of NSI and/or allow for setting tighter constraints on the NSI parameters (see~\cite{Ohlsson:2012kf,Farzan:2017xzy,Dev:2019anc} for reviews). We go beyond these studies and explore the role of NSI on the violation of LGI. We show that the NSI effects could lead to an enhancement in the violation of LGI for appropriate choice of values of these parameters. 
 It would be worthwhile to  explore the impact of NSI in neutrino oscillations via other measures of quantumness such as quantum witness or contextuality~\cite{PhysRevLett.113.140401,PhysRevLett.116.120404,Ban2019,PhysRevA.95.022101,Mendoza-Arenas2019}.

It should be noted that in order to measure the LGI parameter, conventionally, 
  one needs a minimum of three time measurements (for $K_3$). For higher order LGI parameters ($K_n$), we will need four and more such measurements.  
   This means that one requires at least three baselines with identical detection possibilities to infer the simplest of LGI parameters, 
   $K_3$. However, it is practically impossible to realize the three baseline measurement experimentally.   
The value of $\Delta L \simeq 1200$ km where large enhancement in $K_4$ is found does not imply a fixed baseline experiment and should not be confused with any particular experiment.

 A  way to tackle this problem has been put forth by Formaggio et al for two flavour case~\cite{Formaggio:2016cuh}.
 The authors used the fact that in the  phase factor one has two experimental handles - 
 one is the $L$ and other one is the $E$ which can be independently tuned. 
  One can mimic the change in $L$ by a corresponding change in $E$.  This is how the authors performed a test of LGI using data from MINOS experiment with $L=735$ km, 
  by selecting various energies $E_a$ for measurements 
  such that the phases obeyed a certain sum rule. Thus,  using similar approach, if we can observe enhanced violation in data from a fixed baseline experiment such as DUNE, then that will be indicative of  the presence of new physics.

In the present work, we take $| \nu_e \rangle$ as the initial state and discuss the role of current unknowns in neutrino oscillation physics as well as role of antineutrinos in our inferences about LGI violation. We then go on to investigate the impact of NSI on the violation of LGI. In the existing studies related to testing LGI in neutrino oscillations, different initial states have been employed. For instance, $| \nu_\mu \rangle$ was used as an initial state in studies pertaining to LGI for two and three flavour neutrino oscillations~\cite{Naikoo:2017fos,Naikoo:2019eec,Naikoo:2019gme} as well as in establishing $6\sigma$ evidence for violation of LGI in two flavour context using data from MINOS experiment~\cite{Formaggio:2016cuh}. $| \nu_e \rangle$ was used as an initial state in~\cite{Gangopadhyay:2013aha,Gangopadhyay:2017nsn}. While it is found that LGI is violated in all these studies  irrespective of the source type, it would be worthwhile to carry out  a detailed study with different possible source types and investigate their impact on the violation of LGI.

There are discussions on the possibility to manipulate neutrinos for the purpose of communications, such as galactic neutrino communication~\cite{Learned:2008gr} and sabmarine neutrino communication~\cite{Huber:2009kx}. Stancil et al. reported on the performance of a low-rate communications link established using the NuMI beam line and the MINERvA detector  illustrating the feasibility of using neutrino beams to provide low-rate communications link~\cite{Stancil:2012yc}. Synchronized neutrino communications over intergalactic distances have been studied in~\cite{Santos:2020xcn}. Of course, the scales of these studies is  completely different   and averaging of neutrino oscillations will have different manifestations on questions such as LGI and quantum coherence. Temporal correlations of the LGI kind might give additional handle to extend these studies.

\section*{Acknowledgements} 

We acknowledge useful discussions with Ipsika Mohanty and Animesh Sinha Roy during various stages of this work. The use of HPC cluster at SPS, JNU funded by DST-FIST is acknowledged. SS acknowledges financial support in the form of fellowship from University Grants Commission. 
PM would like to acknowledge funding from University Grants Commission under UPE II at JNU and Department of Science and Technology under DST-PURSE at JNU. The work of PM is partially supported by the European Union's Horizon 2020 research and innovation programme under the Marie Skodowska-Curie grant agreement No 690575 and 674896.

\bibliography{references_2021}

\end{document}